\documentclass[a4paper]{jpconf} 
\usepackage{graphicx}

\begin{document}

{\center \jpcs}

\newcommand{\be}{\begin{equation}}
\newcommand{\ee}{\end{equation}}
\newcommand{\bea}{\begin{eqnarray}}
\newcommand{\eea}{\end{eqnarray}}
\def \nn{\nonumber}
\newcommand{\MS}{\ensuremath{\overline{\mbox{MS}}}}

\title{{The double radiative $\bar{B}\to X_s\gamma \gamma$ decay\\
at $O(\alpha_{s})$ in QCD}}

\author{Ahmet Kokulu}

\address{\it Department of Mathematical Sciences, University of Liverpool, L69 3BX Liverpool, \\ United Kingdom}

\ead{akokulu@liverpool.ac.uk}

\begin{abstract}
In these proceedings, we briefly review the individual interference contribution of the electromagnetic dipole operator $\mathcal{O}_{7}$ to the double differential decay width $d\Gamma_{77}/(ds_1 \, ds_2)$ for the process $\bar{B} \to X_s \gamma \gamma$ at $O(\alpha_s)$ in QCD, which is based on our work in \cite{Asatrian:2011ta}. We define two kinematical variables $s_1$ and $s_2$ as $s_i=(p_b - q_i)^2/m_b^2$, where $p_b$, $q_1$, $q_2$ are the momenta of b-quark and two photons. While the (renormalized) virtual corrections are worked out exactly for a certain range of $s_1$ and $s_2$, we retained in the gluon bremsstrahlung process only the leading power w.r.t. the (normalized) hadronic mass $s_3=(p_b-q_1-q_2)^2/m_b^2$ in the underlying triple differential decay width $d\Gamma_{77}/(ds_1 ds_2 ds_3)$. We found that the double differential decay width, based on this approximation, is free of infrared- and collinear singularities when summing up the virtual- and real-radiation corrections, while this was not the case when keeping all powers in $s_3$ in the gluon bremsstrahlung process due to the configurations allowing collinear photon emission from the (massless) $s$-quark. Lastly, we compare our analytical results with those obtained in a recently extended work \cite{Asatrian:2014mwa}, where a non-zero strange quark mass was introduced to regulate the collinear photon configurations.
\end{abstract}

\section{Introduction}

Inclusive rare $B$-meson decays provide a crucial place to probe new physics indirectly. In the Standard Model (SM) all these processes 
proceed through loop diagrams (due to Glashow-Iliopoulos-Maiani mechanism) and thus are relatively suppressed. In the extensions 
of the SM the contributions stemming from the diagrams with ``new'' 
particles in the loops can be comparable or even larger than the contribution from 
the SM. Thus getting experimental information on rare decays puts stringent 
constraints on the extensions of the SM or can even lead to a  
disagreement with the SM predictions, providing evidence for some ``new physics''.

To make a rigorous comparison between experiment and theory, precise
SM calculations for the (differential) decay rates are mandatory. While the
branching ratios for $\bar{B} \to X_s \gamma$ \cite{Misiak:2006zs}
and $\bar{B} \to X_s \ell^+
\ell^-$ are known today even to
next-to-next-to-leading logarithmic (NNLL) precision (for reviews, see
\cite{Hurth:2010tk,Buras:2011we}),
other branching ratios, like the one for $\bar{B} \to X_s \gamma
\gamma$ discussed in these proceedings, has been calculated before to leading logarithmic
(LL) precision in the SM by several groups \cite{Simma:1990nr,Reina:1996up,Reina:1997my,Cao:2001uj} and only recently a first step towards next-to-leading-logarithmic (NLL) precision was presented by us in \cite{Asatrian:2011ta}.
In contrast to $\bar{B} \to X_s
\gamma$, the current-current operator ${\cal O}_2$ has a non-vanishing matrix
element for $b \to s \gamma \gamma$ at order $\alpha_s^0$ precision, 
leading to an interesting interference pattern with the contributions associated
with the electromagnetic dipole operator ${\cal O}_7$ already at LL
precision. As a consequence, potential new physics should be clearly visible
not only in the total branching ratio, but also in the 
differential distributions.

As the process $\bar{B} \to X_s \gamma \gamma$ is expected to be measured at
the planned Super $B$-factory in Japan ({\bf SuperKEKB}), it is necessary
to calculate the differential distributions to NLL precision in the
SM, in order to
fully exploit its potential concerning new physics.

While the Wilson coefficients $C_i(\mu)$ (appearing in the definition of the effective Hamiltonian)
are known to sufficient precision at the low scale $\mu \sim m_b$
since a long time (see e.g. the reviews \cite{Hurth:2010tk,Buras:2011we}
and references therein), the matrix elements 
$\langle s \gamma \gamma|{\cal  O}_i|b\rangle$ and 
$\langle s \gamma \gamma \, g|{\cal  O}_i|b\rangle$, 
which in a NLL calculation are needed to order
$g_s^2$ and $g_s$, respectively, are not fully known yet. To calculate the
$({\cal O}_i,{\cal O}_j)$-interference contributions to the
differential distributions at order
$\alpha_s$ is in many respects of similar complexity as the
calculation of the photon energy spectrum in $\bar{B} \to X_s \gamma$ 
at order $\alpha_s^2$
needed for the NNLL computation. As a first step in this NLL enterprise, we
derived in our paper \cite{Asatrian:2011ta}, the $O(\alpha_s)$
corrections 
to the $({\cal O}_7,{\cal O}_7)$-interference contribution to the double 
differential decay width $d\Gamma/(ds_1 ds_2)$ at the partonic level. We defined the variables $s_1$
and $s_2$ as $s_i=(p_b-q_i)^2/m_b^2$, where $p_b$
and $q_i$ denote the four-momenta of the $b$-quark and the two
photons, respectively.

At order $\alpha_s$
there are contributions to $d\Gamma_{77}/(ds_1 ds_2)$ with three
particles 
($s$-quark and two photons)
and four particles ($s$-quark, two photons and a gluon) in the final state.
These contributions correspond to specific cuts of the $b$-quarks
self-energy at order $\alpha^2 \times \alpha_s$, involving twice the
operator ${\cal O}_7$. As there are additional cuts, which contain for
example only one photon, our observable cannot be obtained using the
optical theorem, i.e., by taking the absorptive part of the $b$-quark
self-energy at three-loop. We therefore calculated the mentioned 
contributions with three and four particles in the final state individually.

We work out the QCD corrections
to the double differential decay width 
in the kinematical range (see the left frame of Fig. \ref{fig:results})
\[
0 < s_1 < 1 \quad ; \quad 0 < s_2 < 1-s_1 \, .
\]

Concerning the virtual corrections, all singularities (after
ultra-violet renormalization) are due to  {\bf soft gluon} exchanges
and/or  {\bf collinear gluon} exchanges involving the $s$-quark. Concerning the
bremsstrahlung corrections (restricted to the same range of $s_1$ and
$s_2$), there are in addition kinematical situations where {\bf collinear photons}
are emitted from the $s$-quark. The corresponding singularities did not
cancel when combined with the virtual corrections. We found, however, that there are no
singularities associated with collinear photon emission in the double
differential decay width when only retaining
the leading power w.r.t to the (normalized) hadronic mass
$s_3=(p_b - q_1 - q_2)^2/m_b^2$ in the underlying triple differential distribution
$d\Gamma_{77}/(ds_1 ds_2 ds_3)$. The results of our paper \cite{Asatrian:2011ta} were
obtained within this ``approximation''.

\section{The leading and the $O(\alpha_{s})$ results for the decay width}\label{sec:combination}
In $d=4$ dimensions, the leading-order spectrum (in our restricted
phase-space) is given by
\begin{eqnarray}
&&\frac{d\Gamma_{77}^{(0)}}{ds_1 \, ds_2} = \frac{\alpha^2 \, \bar{m}_b^2(\mu) \, m_b^3 \, |C_{7,eff}(\mu)|^2 \, G_F^2 \,
  |V_{tb} V_{ts}^*|^2 \,  Q_d^2}{1024 \, \pi^5} \, 
  \frac{(1-s_1-s_2)}{(1-s_1)^2 s_1 (1-s_2)^2 s_2} \, r_0 \, .
\label{treezero}
\end{eqnarray}
where
\begin{eqnarray}
\nonumber r_0&=&-48 s_2^3 s_1^3+96 s_2^2 s_1^3-56 s_2 s_1^3+8
   s_1^3+96 s_2^3 s_1^2-192 s_2^2 s_1^2+112
   s_2 s_1^2-56 s_2^3 s_1+
   \\&&\nonumber 112 s_2^2 s_1-96 s_2
   s_1+8 s_1+8 s_2^3+8 s_2
    \end{eqnarray}

Adding the renormalized virtual corrections and the bremsstrahlung corrections, the complete order $\alpha_s$ correction to the
double differential decay width
$d\Gamma_{77}/(ds_1 \, ds_2)$ reads
\begin{eqnarray}
  \frac{d\Gamma_{77}^{(1)}}{ds_1 \, ds_2} = 
\frac{\alpha^2 \, \bar{m}_b^2(\mu) \, m_{b}^3 \, |C_{7,eff}(\mu)|^2 \, G_F^2 \,
  |V_{tb} V_{ts}^*|^2 \,  Q_d^2}{1024 \, \pi^5}        
  \times \frac{\alpha_s}{4\pi} \, C_F  \, \left[
\frac{-4 \, r_0 \, (1-s_1-s_2)}{(1-s_1)^2 \, s_1 \, (1-s_2)^2 \, s_2} \, \log \frac{\mu}{m_b} +f
\right],
\label{total}
\end{eqnarray}
where the explicit expression for $f$ can be found in \cite{Asatrian:2011ta}. 

The order $\alpha_s$ correction $d\Gamma_{77}^{(1)}/(ds_1 ds_2)$ in
Eq. (\ref{total}) to the double differential decay width
for $b \to X_s \gamma \gamma$ was the
main result of our paper \cite{Asatrian:2011ta}.

\section{Numerical illustrations I}\label{sec:numerics}
\begin{figure}[h]
\centering{
\includegraphics[width=0.24\textwidth]{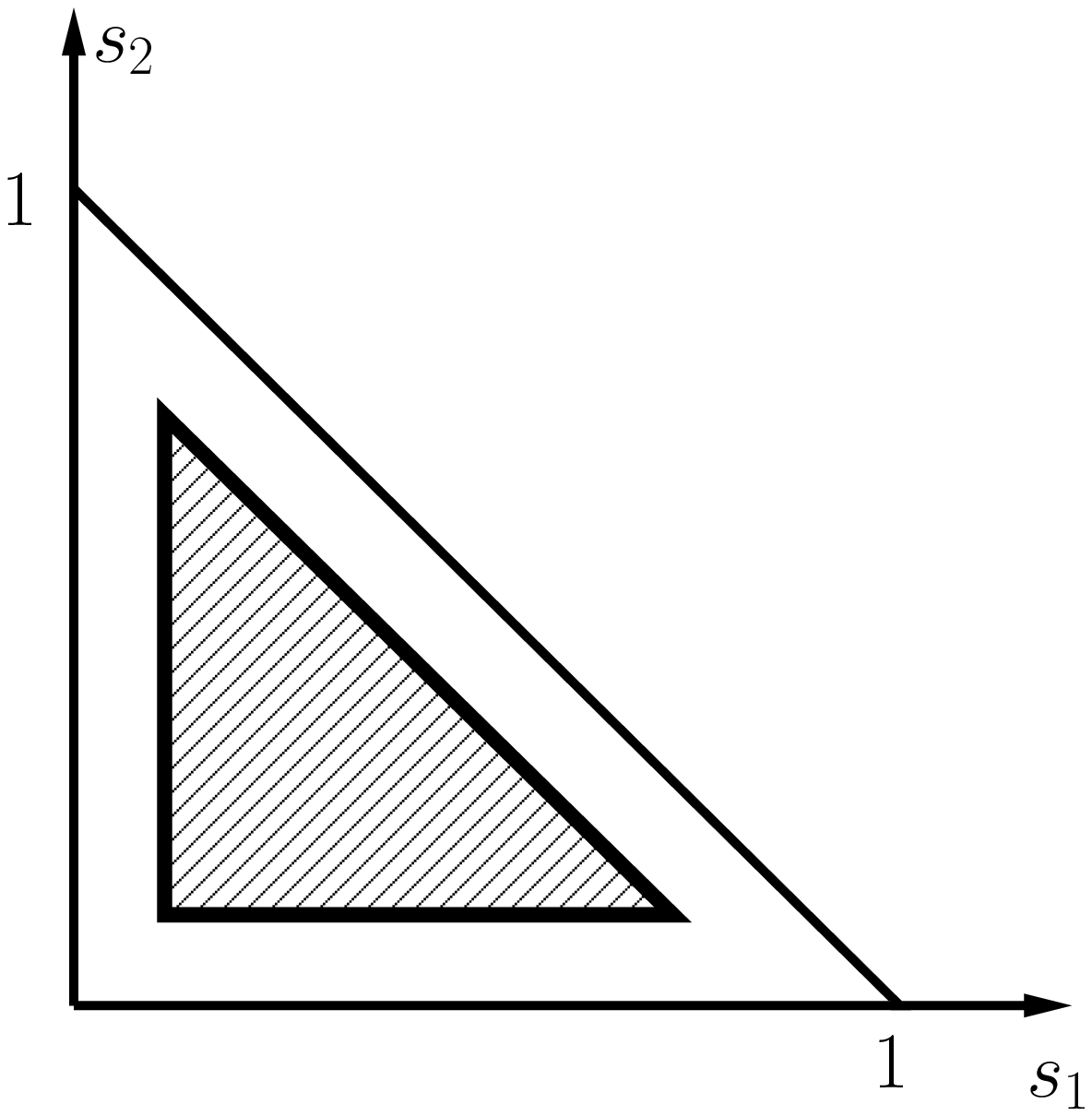}
\hspace{0.9cm}
\includegraphics[width=0.34\textwidth]{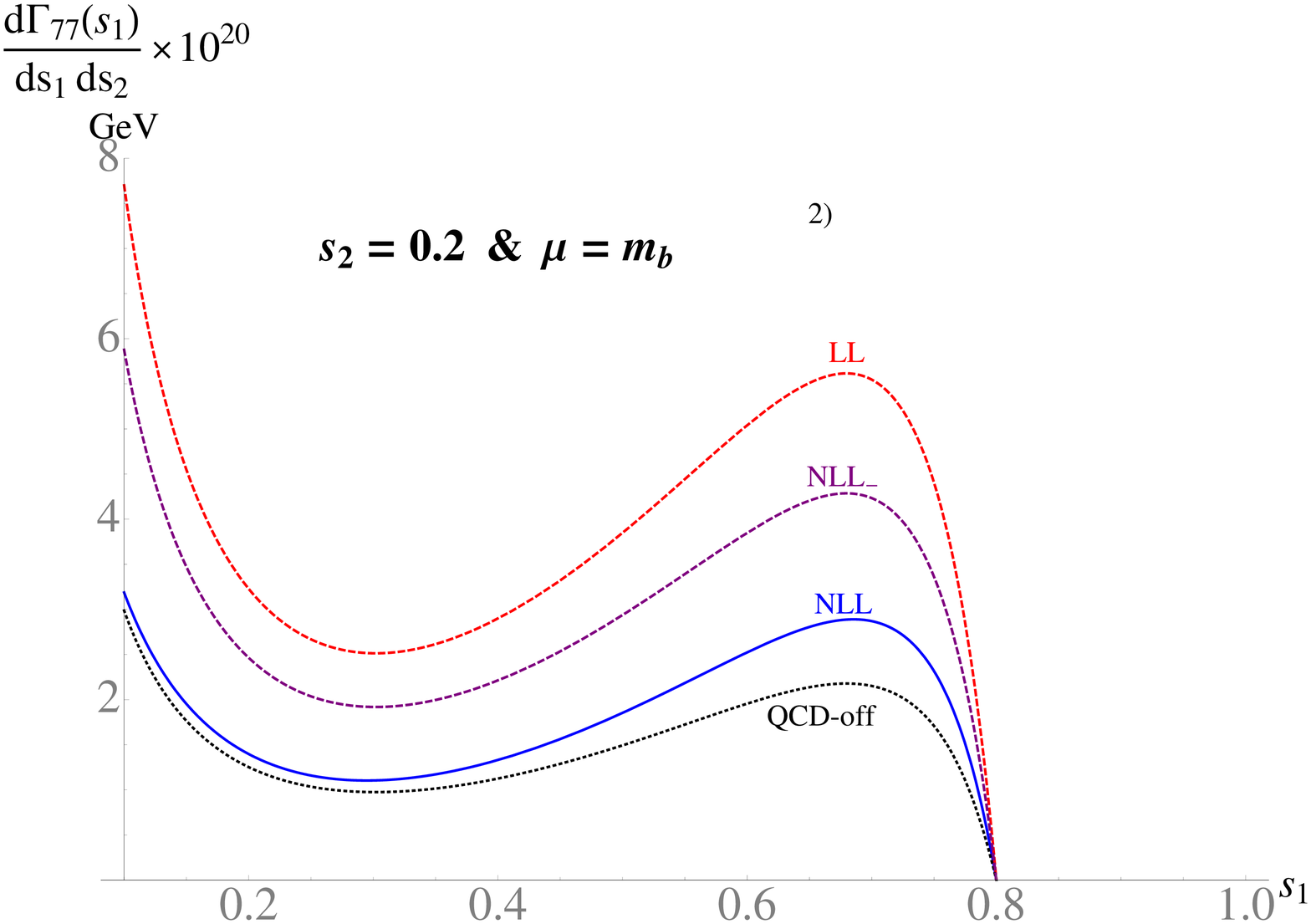} 
\vspace{0.2cm}
\includegraphics[width=0.32\textwidth]{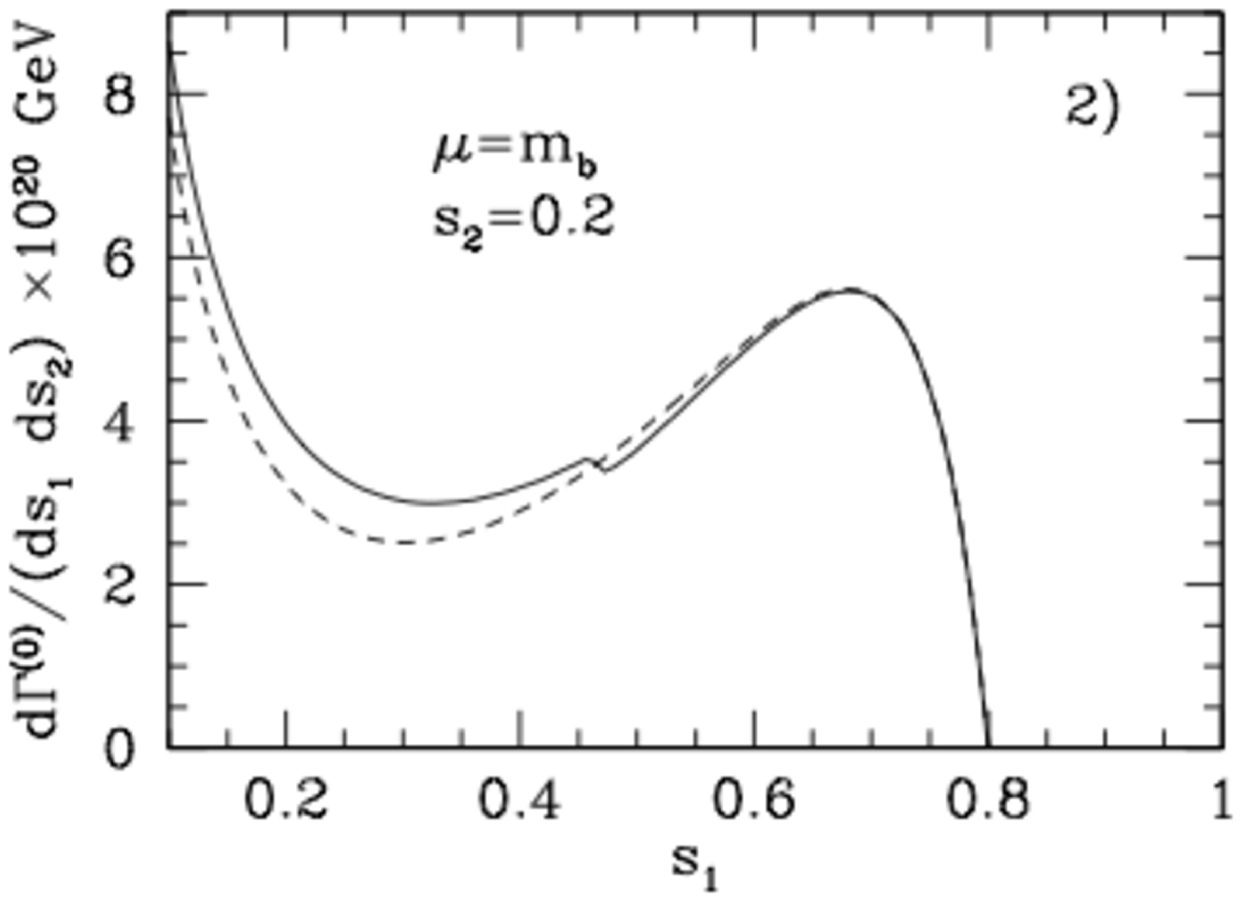}  }  
\caption{\footnotesize{ {\bf Left:} The relevant phase-space region for $s_{1}$ and $s_{2}$ used in this paper. {\bf Middle:} Double differential decay width, based on the operator ${\cal O}_{7}$ only, as a function of $s_1$ for $s_{2}$ fixed at $s_2=0.2$ and $\mu=m_{b}$. 
The dotted(black), the short-dashed(red) and the solid line(blue) shows the result 
when neglecting QCD-effects, the LL and the NLL result,
respectively. The long-dashed line(purple) represents the (partial) NLL result
in which the virtual- and bremsstrahlung corrections worked out in our
paper \cite{Asatrian:2011ta} are switched off. {\bf Right:} Plot taken from Ref.~\cite{Asatrian:2014mwa} illustrating the LL decay width as a function of $s_1$ for $s_{2}$ fixed at $s_2=0.2$ and $\mu=m_{b}$. In this frame, the dashed line shows the result when only the (${\cal O}_7$, ${\cal O}_7$) interference is taken into account, while the solid line shows the result when all contributions associated with ${\cal O}_1$, ${\cal O}_2$ and ${\cal O}_7$ are included.}}
\label{fig:results}
\end{figure}

In our procedure the NLL corrections have three sources: 
(a) $\alpha_s$ corrections to the Wilson coefficient $C_{7,eff}(\mu)$,
(b) expressing $\bar{m}_b(\mu)$ in terms of the pole mass $m_b$ and
(c) virtual- and real- order $\alpha_s$ corrections to the matrix
elements. To illustrate the effect of source (c), which was worked out
for the first time in our paper \cite{Asatrian:2011ta} (see also Ref. \cite{Kokulu:2012tw} for a brief review), we show in the middle frame of Fig. \ref{fig:results}
(by the long-dashed line) the (partial) NLL result in which 
source (c) is switched off. 
We conclude that the effect (c) is roughly of equal
importance as the combined effects of (a) and (b). 

For completeness we
show in the middle frame of this figure (by the dotted line) also the result when QCD
is completely switched off, which amounts to put $\mu=m_W$ in the LL
result. Furthermore, the right frame of Fig. \ref{fig:results} illustrates the LL spectrums for the cases when taking into account the (${\cal O}_{7}$, ${\cal O}_{7}$) interference contribution only (dashed curve) and also when all contributions associated with ${\cal O}_{1}$, ${\cal O}_{2}$ and ${\cal O}_{7}$ are considered (solid curve).

From Fig. \ref{fig:results} we see that the NLL results are
substantially smaller (typically by $50\%$ or slightly more) than
those at LL precision, which was also the case when choosing other
values for $s_2$. 

In the numerical discussion above, 
we observed the following:
Generally speaking, NLL corrections are not small, when taking
into account the full range of $\mu$, i.e., $m_b/2<\mu<2 \, m_b$.

\section{Numerical illustrations II}\label{sec:numerics2}
\begin{figure}[h]
\centering{
\includegraphics[width=0.38\textwidth]{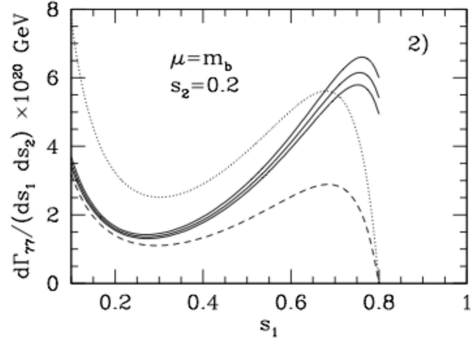} }
\caption{\footnotesize{ Plot taken from Ref.~\cite{Asatrian:2014mwa} illustrating the double differential decay width, based on the operator ${\cal O}_7$ only, as a function of $s_1$ for $s_2$ fixed at $s_2 = 0.2$ and $\mu=m_{b}$. The dotted, the dashed and the solid lines show the LL result, the NLL when only retaining leading power terms as in Ref. \cite{Asatrian:2011ta} and the full NLL result of the recent paper \cite{Asatrian:2014mwa}, respectively. Among the three solid lines, the highest, middle and lowest curve correspond to setting $m_s = 400$ MeV, $m_s = 500$ MeV and $m_s = 600$ MeV, respectively.            }}
\label{fig:resultsGreub14}
\end{figure}

In this section we briefly compare the result of the present work \cite{Asatrian:2011ta} with those obtained in a recent analysis \cite{Asatrian:2014mwa}. In the work of \cite{Asatrian:2014mwa}, a non-zero mass for the $s$-quark was introduced in order to regulate the collinear photon emissions from the $s$-quark, retaining all powers w.r.t. hadronic mass $s_{3}$. As can be seen from Fig. \ref{fig:resultsGreub14}, in either case (work \cite{Asatrian:2011ta} or work \cite{Asatrian:2014mwa}), the NLL QCD contributions to the decay width are large in general.

\section{Summary}\label{sec:summary}
In these proceedings we reviewed the calculation of the set of $O(\alpha_s)$
corrections to the observable $\bar{B} \to X_s \gamma \gamma$
originating from diagrams involving ${\cal O}_7$.
To perform this calculation, it was necessary to work out diagrams
with three particles ($s$-quark and two photons) and four
particles ($s$-quark, two photons and a gluon) in the final state.
From the technical point of view, the calculation was made possible
by the use of the Laporta Algorithm to identify the
needed master integrals and by applying the differential equation method to
solve the master integrals.
When calculating the bremsstrahlung corrections, we took into account
only terms proportional to the leading power of the hadronic mass.
We found that the infrared and collinear singularities cancel when
combining the above mentioned approximated version of bremsstrahlung
corrections with the virtual corrections.
The numerical impact of the NLL corrections is found to be large: for
$d\Gamma_{77}/(ds_1 \, ds_2)$ the NLL result is approximately
50\% smaller than the LL prediction. 

Moreover, we compared our present results with those obtained in a recent paper \cite{Asatrian:2014mwa}, where a massive strange quark was considered to regulate the collinear photon (gluon) configurations, retaining all powers w.r.t. the hadronic mass.

\ack
I wish to thank the organisers of the conference TROIA'14 for their efforts to make it very pleasant and for the possibility to present these results. \\[0.3cm]

\bibliography{kokulu_TROIA14}

\end{document}